\documentclass[longbibliography,reprint,2,amsmath,amssymb,aps]{revtex4-1}

\usepackage{graphicx}
\usepackage{dcolumn}
\usepackage{bm}
\usepackage{natbib}
\usepackage[table]{xcolor}
\usepackage{tabularx}
\usepackage{array}
\usepackage{colortbl}
\usepackage{appendix}
\usepackage{amsmath}

\begin{document}
\preprint{APS/123-QE}

\title{Optimal probe states for single-mode quantum target detection in arbitrary object reflectivity}
\author{Wei-Ming Chen$^2$, Pin-Ju Tsai$^{1,3}$}
\email{tpinju@ncu.edu.tw}
\affiliation{$^{1}$Department of Optics and Photonics, National Central University, Taoyuan City 320317, Taiwan
}
\affiliation{$^{2}$Department of Physics, National Central University, Taoyuan City 320317, Taiwan
}
\affiliation{$^{3}$Center for Quantum Science and Technology, National Central University, Taoyuan City 320317, Taiwan
}
\date{\today}
\begin{abstract}

Quantum target detection (QTD) utilizes nonclassical resources to enable radar-like detection for identifying reflecting objects in lossy and noisy environments, surpassing the detection performance achieved by classical methods. To fully exploit the quantum advantage in QTD, determining the optimal probe states (OPSs) across various detection parameters and gaining a deeper understanding of their characteristics are crucial. In this study, we employ optimization algorithms to identify the single-mode continuous-variable OPSs for entire range of target reflectivity. Our findings suggest that OPSs are non-Gaussian states in most reflectivity scenarios, with exceptions under specific conditions. Furthermore, we provide a comprehensive physical interpretation of the observed phenomena. This study offers a tool for identifying OPSs along with a clear physical interpretation. It also contributes to further advancements towards optimal multi-mode QTD, which holds the potential for broad applications in quantum sensing and metrology.

\end{abstract}
\maketitle
\section{Introduction}
Quantum target detection (QTD) is a critical task within the field of quantum sensing. In the context of QTD, a reflecting target is situated amidst a lossy and noisy environment, like typical radar detection scenarios. The probe emits a quantum signal to the target area, receives the reflected signal, and further analyzes its quantum nature to ascertain the presence of the target \cite{pirandola2018advances,giovannetti2011advances,polino2020photonic}. To enhance the effectiveness of QTD, Lloyd initially introduced discrete variable (DV) entanglement into QTD, known as quantum illumination (QI) \cite{lloyd2008enhanced,shapiro2020quantum}. Subsequently, the concept of QI was applied to various types of quantum states as probes, including Gaussian states in continuous-variable (CV) systems \cite{guha2009gaussian,pirandola2008computable,shapiro2009quantum,tan2008quantum,nair2020fundamental}, and even non-Gaussian states \cite{zhang2014quantum,sanz2017quantum}. In 2013, the first experimental realization of quantum illumination was presented \cite{PhysRevLett.110.153603} based on the protocol proposed by Lloyd and S. Tan et al. \cite{lloyd2008enhanced,tan2008quantum}. Furthermore, a series of experimental demonstrations of QTD based on QI was presented in the optical frequency region \cite{zhang2013entanglement,zhang2015entanglement} and even in the microwave region \cite{barzanjeh2020microwave,assouly2023quantum,chang2019quantum,luong2019receiver,bourassa2020progress,livreri2022microwave,livreri2023microwave,hosseiny2022engineered}, laying the foundation for the implementation of type-III quantum lidar/radar \cite{lanzagorta2012quantum,burdge2009quantum,torrome2023advances,karsa2023quantum,torrome2020introduction,al2020quantum}. In more practical application scenarios, quantum radar systems must advance beyond previous quantum illumination protocols, which merely identify the presence of a target within a specific space. Recently, there has been further discussion on the three-dimensional localization of targets by quantum radar. Positioning within three-dimensional space by utilizing the protocol of Gaussian beam entangled photons in the frequency domain was first proposed in \cite{maccone2023gaussian}, which further demonstrates a $\sqrt{N}$ times quantum enhancement over the unentangled case when $N$ photons are used. Moreover, achieving more precise azimuth resolution of targets through the use of dual-receiver quantum radar protocols has also been discussed \cite{wu2023entanglement}.

In the aforementioned work, these studies are all based on a known probe quantum state as the foundation for research. The discussion revolves around the quantum advantage of this quantum state under stringent probing conditions (high loss and noisy environment). For example, in \cite{shapiro2009quantum}, a two-mode squeezed vacuum (TMSV) is discussed as the probe quantum state and achieves a quantum advantage of 6 dB. However, an intriguing question arises: are these discussed quantum states the optimal probe states (OPSs)? Is achieving a higher advantage using other types of quantum states under the same probing conditions possible? 

In DV systems, the issue of OPSs is addressed in \cite{yung2020one}, where the maximally entangled state is identified as the OPS. Subsequent experiments demonstrate that using the maximally entangled state as the probe state can approach the theoretical limit of the Helstrom bound \cite{xu2021experimental}. In CV systems, much research focuses on verifying the OPSs under conditions of low reflection targets. In \cite{bradshaw2021optimal}, Mark et al. employ mathematical methods (Lagrange multiplier) to investigate the OPSs for quantum target detection under extreme conditions (high noise and low target reflectivity or high-loss channels) for both single-mode (without entanglement) and two-mode (involving entanglement) detection scenarios. They conclude that under single-mode and extreme detection conditions, the OPS approximates a coherent state, while for two-mode scenarios, the OPS corresponds to TMSV, which is consistent with the results presented in \cite{de2018minimum}. Furthermore, in the case of given probe states, optimizing the parameters of the probe state's properties is also an essential topic for QTD. In \cite{spedalieri2021optimal}, the study focuses on using displaced squeezed states as probe states and investigates the joint optimal squeezing and displacement parameters of these states under various target reflectivities. It further demonstrates that these states can outperform coherent states with the same mean number of input photons.

Although studies on OPSs have provided solutions under strict conditions (such as extremely high losses or low target reflectivity), there has been no clear discussion on CV OPSs for targets with finite or even high reflectivity. These non-extreme conditions more accurately reflect real-world detection scenarios. Moreover, the study of the arbitrary reflectivity of objects could also be analogous to various detection tasks. For example, when dealing with targets exhibiting high reflectivity, the situation resembles scenarios where the optical field is reflected back and detected under conditions of low loss. This implies weak interaction between the target and the probing light, a condition commonly encountered in measurements involving biological samples \cite{taylor2016quantum,PhysRevResearch.2.033243,whittaker2017absorption,degen2017quantum,petrini2020quantum}.

In this study, to comprehend OPSs under global reflectivity, we utilize optimization algorithms to identify the single-mode OPS that minimizes the error probability of QTD under arbitrary detection conditions. We demonstrate that the OPSs we identified exhibit higher identification performance than coherent state probes in QTD across arbitrary detection conditions. Drawing from the observed behavior of OPS under different conditions, we also delve into OPS from various physical perspectives to offer reasonable interpretations of its behavior. Our analysis shows that in most cases, single-mode OPSs are non-Gaussian, but under specific background noise and target reflectivity conditions, OPS can revert to coherent states. Furthermore, an important finding is that the probe-obtained information for distinguishing target absence or presence can be determined by the probe state's phase or photon number distribution. This insight allows us to distinguish OPSs into two regimes based on target reflectivity: those dominated by phase-squeezed states and those dominated by photon number squeezed states (PNSSs), confirming the types of OPSs in given detection conditions. Our work provides a comprehensive understanding of OPSs under arbitrary target reflectivity conditions, offering insights for optimizing QTD techniques. The developed approach can be extended to explore higher mode quantum states for optimal probe states with quantum entanglement.

The paper is structured as follows: In Sec.\ref{secII}, we present the theoretical model of quantum target detection, covering aspects such as error probability estimation, beam splitter (BS) model, and optimal probe states. Subsequently, in Sec.\ref{secIII}, we provide numerical results and discuss the underlying physical concepts. To present our findings systematically, we initially examine an athermal environment (noise-free environment) in Sec.\ref{secIII}.A. Building upon the insights gained in this section, we then extend our discussion to a more general scenario of a thermal environment in Sec.\ref{secIII}.B. In Sec.\ref{secIV}, we summarize the results of this study. The optimal algorithm and the details for calculation are described in the Appendix. 

\begin{figure*}[t]
\centering
\includegraphics[width=0.92\textwidth]{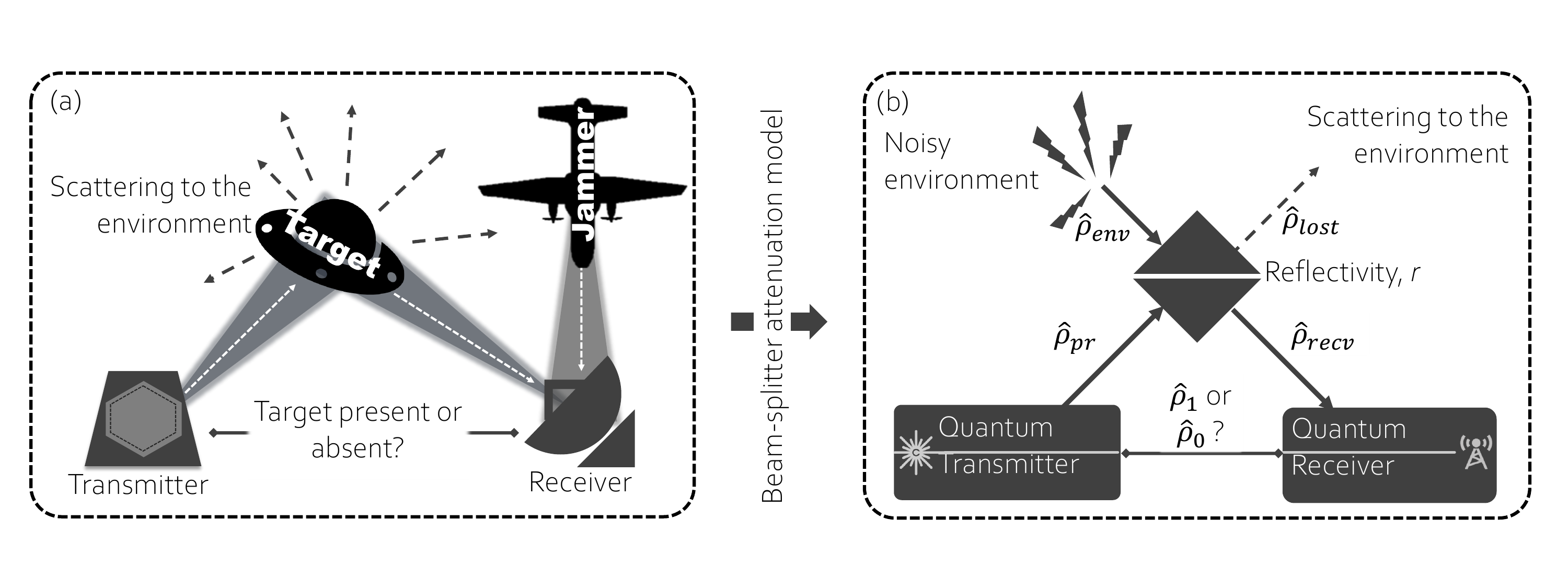}
\caption{(a) Schematic diagram of radar detection scenario. (b) QTD with the beam-splitter (BS) model. The quantum transmitter sends a probe quantum state $\hat{\rho}_{pr}$ to the BS, which can be used to analogy the lossy channel and un-unitary reflected object. To simulate the noisy environment, the thermal state $\hat{\rho}_{env}$ is coupled into the BS from another input mode and mixed with the probe state. Then, the quantum receiver finally receives $\hat{\rho}_{recv}$ to process the hypotheses testing for target detection. The state $\hat{\rho}_{lost}= \text{Tr}_{recv}[\hat{\rho}^{out}]$ is the quantum state leak to the environment, which will not be detected.}
\label{Fig1}
\end{figure*}

\section{Theoretical model} \label{secII}

Quantum target detection is a scenario similar to common radar detection, as illustrated in Fig.\ref{Fig1} (a). In this scenario, a quantum transmitter emits quantum light to illuminate the target within a noisy environment, and a quantum receiver is employed to capture the reflected signal. The objective of the detector is to distinguish between two hypotheses, $H_{0,1}$, in a binary test. 
\begin{itemize}
\item[$H_0$:]The target is absent, and the receiver obtains the quantum state $\hat{\rho}_0$ which is all occupied by noise states.
\item[$H_1$:]The target is present, and parts of the probe state mixing with noise states will be detected by the receiver and identify the state as $\hat{\rho}_1$.
\end{itemize}
In this sense, two hypotheses $H_{0,1}$ are represented by two different quantum states, $\hat{\rho}_{0,1}$, respectively. Thus, briefly speaking, the hypotheses testing task is a problem of quantum state discrimination. Due to the probable harsh detection scenario, however, the quantum states $\hat{\rho}_{0}$ and $\hat{\rho}_{1}$ may be similar to each other and further make some error in distinguishing the hypotheses $H_{0,1}$. To estimate the error, the Helstrom bound \cite{helstrom1969quantum} offers a theoretical limit on the minimum error probability when distinguishing between the two states in a single measurement of $\hat{\rho}_{0}$ and $\hat{\rho}_{1}$,
\begin{equation}
P_{err} = \frac{1-||p_0\hat{\rho}_0-p_1\hat{\rho}_1||_{1}}{2},
\label{P_err}
\end{equation}
where $||M||_{1}\equiv tr\sqrt{M^\dagger M}$ is trace norm of matrix $M$, $p_0$ is prior probabilities of $H_0$ and $p_1=1-p_0$ for $H_1$. Note that Eq.\ref{P_err} holds only when employing the optimal receiver. If a non-optimal receiver is utilized, the error probability behavior may differ. In practice, the choice of quantum receivers in a quantum sensing system depends on the quantum state of the light or the encoding dimension. This selection aims to extract the quantum state of the returned probing light field with utmost accuracy, thereby minimizing the error probability. As an example, when utilizing Gaussian states as CV probe states, potential solutions involve the utilization of quantum receivers like homodyne detection\cite{shapiro2009quantum,jo2021quantum}, optical parametric amplifier (OPA), and phase-conjugate receiver \cite{guha2009gaussian,zhuang2017optimum}. On the other hand, for DV probe states, extensive research has been conducted on using the measurement of Bell states as an optimal quantum receiver \cite{yung2020one,xu2021experimental}. However, within the context of identifying the optimal probing quantum states discussed in this study, it may not necessarily be a Gaussian state. In fact, it can be any quantum state for optimizing sensing tasks. Therefore, we assume that the quantum receivers are optimized for all the probe states under consideration.

To estimate the error probability $P_{err}$, the states $\hat{\rho}_{0}$ and $\hat{\rho}_{1}$ must be calculated. Here, we utilize a BS model to analogy the lossy and noisy detection scenario, as shown in Fig.\ref{Fig1} (b), and further calculate the states $\hat{\rho}_{0}$ and $\hat{\rho}_{1}$. In the BS model, a quantum transmitter inputs a quantum state of $\hat{\rho}_{pr}$ as a probe to a BS as the target with a reflectivity $r$. The noisy environment is simulated by coupling into another mode of BS (see Fig.\ref{Fig1}). To facilitate comparison with the majority of research on quantum target detection \cite{guha2009gaussian,tan2008quantum,nair2020fundamental,zhang2014quantum,bradshaw2021optimal}, in this case, the environmental noise quantum state is considered a mixed thermal state $\hat{\rho}_{env}$. The thermal state is given by
\begin{equation}
\hat{\rho}_{env}=\sum_{n=0}^{\infty}\frac{\bar{n}_{env}^n}{\left(\bar{n}_{env}+1\right)^{n+1}}\left|n\right\rangle\left\langle n\right|, 
\end{equation}
where $\bar{n}_{env}$ is the mean photon number of noise. Notice that the state $\hat{\rho}_{env}$ has no off-diagonal term in the density matrix at Fock basis, which means the state has no phase information. For the probe state $\hat{\rho}_{pr}$, since the phase-less nature of $\hat{\rho}_{env}$, the suitable form of the probe state is to consider it as a single-mode pure state $\left|\psi\right\rangle_{pr}$ to maximally the performance for distinguish the $\hat{\rho}_0$ and $\hat{\rho}_1$,
\begin{equation}
\left|\psi\right\rangle_{pr}=\sum_{n=0}^{\infty}c_n\left|n\right\rangle, 
\label{probe}
\end{equation} 
and $\hat{\rho}_{pr}=\left|\psi\right\rangle_{pr}\left\langle\psi\right|$. To evaluate the output quantum state after interacting with the target, we characterize the BS as a quantum process tensor $\mathcal{E}_{BS}$, and further, the density matrix elements of the total output state, $\hat{\rho}^{out}$, can be represented as
\begin{equation}
\rho^{out}_{j_1k_1j_2k_2}=\sum_{m_1,n_1,m_2,n_2\in\mathbb{N}_0}\mathcal{E}_{j_1k_1j_2k_2}^{m_1n_1m_2n_2}\rho^{in}_{m_1n_1m_2n_2},
\end{equation}
where $\rho^{in}_{m_1n_1m_2n_2}$ is the density matrix elements of total input state of $\hat{\rho}^{in}=\hat{\rho}_{pr}\otimes\hat{\rho}_{env}$, and 
\begin{widetext}
\begin{equation}
\begin{aligned}
\mathcal{E}_{j_1k_1j_2k_2}^{m_1n_1m_2n_2}=&\sqrt{\frac{m_{1}!m_{2}!n_{1}!n_{2}!}{j_{1}!j_{2}!k_{1}!k_{2}!}}\sum_{p=0}^{j_{1}}\sum_{q=0}^{k_{1}}\binom{j_{1}}{p}\binom{j_{2}}{m_1-p}\binom{k_{1}}{q}\binom{k_{2}}{n_1-q}\\
&\times\sqrt{1-r^2}^{2p+2q+j_2+k_2-m_1-n_1}(-1)^{j_1+k_1-p-q}\sqrt{r}^{j_1+k_1+m_1+n_1-2p-2q}\delta_{m_1+m_2,j_1+j_2}\delta_{n_1+n_2,k_1+k_2}
\end{aligned}
\end{equation}
\end{widetext}
is the process tensor element of BS in the Fock basis \cite{rahimi2011quantum}. Based on this BS setup, the total output state comprises two modes: the transmitted mode and the reflected mode. Since the quantum receiver only receives the mode reflected from the target, we trace over the lost mode on the total output state as $\hat{\rho}_{recv}=\text{Tr}_{lost}[\hat{\rho}^{out}]$ to obtain the quantum state that finally is received by the quantum receiver. Notice that since the noise state has no defined phase, it will not interfere with the probe states. Under the model, two hypotheses can be easily described. In the case of $H_0$ indicating target absence, we have $\hat{\rho}_{0}= \hat{\rho}_{env}$. For $H_1$ suggesting target presence, we get $\hat{\rho}_{1}=\hat{\rho}_{recv} = \text{Tr}_{lost}[\hat{\rho}^{out}]$. By utilizing the earlier result and Eq.\ref{P_err}, we can further calculate the error probability for a given probe state $\hat{\rho}_{pr}$. 

Now, our attention shifts to determining the OPSs for single-mode quantum target detection. The objective for these specified OPSs is to minimize $P_{err}$ under specific conditions, such as target reflectivity $r$ and mean environmental noise photon $\bar{n}_{env}$. To achieve this, the OPSs must arrange the composition of $c_n$ in Eq.\ref{probe} for minimizing $P_{err}$. However, the composition of $c_n$ is not arbitrary, subject to two important constraints. Firstly, the normalization of quantum states requires that the probe states must obey the rule
\begin{equation}
\begin{aligned}
\sum_n \left|c_n\right|^2=1. 
\label{nor}
\end{aligned}
\end{equation}
Secondly, the mean photon number of probe states should be limited to a finite value; thus, 
\begin{equation}
\begin{aligned}
\sum_n \left|c_n\right|^2 n=\bar{n},
\label{n_bar}
\end{aligned}
\end{equation}
where $\bar{n}$ is the mean photon number of probe states. 

Based on the model above, one can understand that the error probability $P_{err}$ is a function of the variable set of $\{c_n\}$ with the constraints of Eq.\ref{nor} and Eq.\ref{n_bar}. Once a specific set of $\{c_n\}$ which lets the $P_{err}$ become minima, the set of $\{c_n\}$, the corresponding probe states, is called optimal probe states. To find the OPSs, we regard the error probability $P_{err}$ as a target function to perform the Sequential Quadratic Programming (SQP) \cite{boggs1995sequential}. After processing the algorithm (see Appendix), the output set of $\{c_n\}$ or the OPSs are obtained. 

To emphasize the difference between the evolved OPSs and conventional coherent state probing, we first set the initial condition of $\{c_n^{coh}\}$ for the algorithm as a coherent state of 
\begin{equation}
\begin{aligned}
c_n^{coh}=e^{-\bar{n}/2}\frac{\bar{n}^{n/2}}{\sqrt{n!}}.
\label{coh}
\end{aligned}
\end{equation}
Then, we calculate the error probabilities by substituting both the coherent state and OPSs into Eq.\ref{P_err}, respectively, resulting in $P_{err}^{coh}$, representing the error probability when the coherent state is used, and $P_{err}^{opt}$, representing the error probability when optimal probe states are used. To show the advantage of optimal probe states, we compare $P^{coh}_{err}$ and $P^{opt}_{err}$ by introducing a quantum advantage (QA) of
\begin{equation}
\text{QA}({\rm dB})=10\log_{10}\left(\frac{P^{coh}_{err}}{P^{opt}_{err}}\right)
\label{QA}
\end{equation}
in the following discussions. 

\begin{figure*}[t]
\centering
\includegraphics[width=0.92\textwidth]{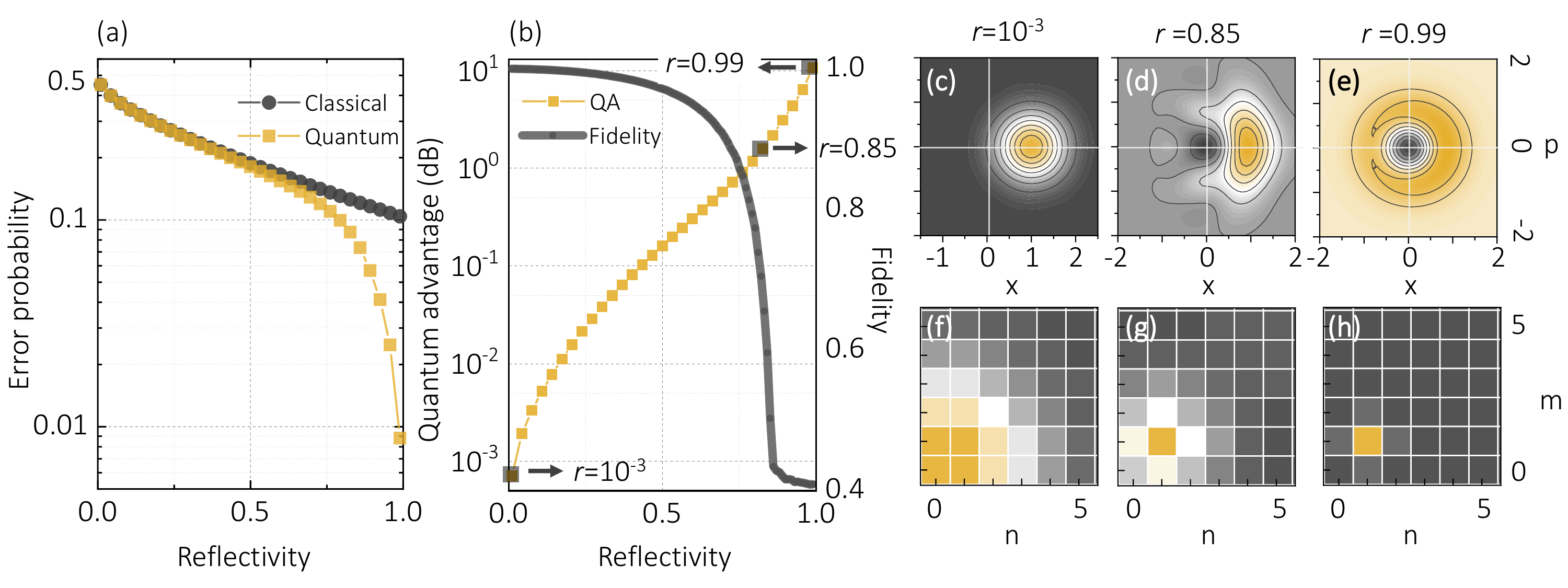}
\caption{(a) Error probability of classical (black) and quantum (yellow) cases. (b) Quantum advantage (yellow) and fidelity (black) of optimal probe states. (c-e) Demonstration of Wigner functions of optimal probe states at a reflectivity of $10^{-3}$, 0.85, and 0.99, respectively. (f-h) Density matrices of optimal probe state at the same conditions as above. In all cases, the mean photon number is set at $\bar{n}=1$.}
\label{Fig2}
\end{figure*}

\section{Results and discussion}\label{secIII}

Utilizing the theoretical tools introduced earlier, we will analyze the properties of OPSs under various scenarios and conditions in this section. Specifically, we will examine a scenario where the probe possesses no prior knowledge of the target, denoted as $p_0=p_1=0.5$. In this case, the error probability (Eq.\ref{P_err}) simplifies to $(1-\frac{1}{2}||\hat{\rho}_0-\hat{\rho}_1||_{1})/2$, representing the most generic detection scenario.

To gain a clear physical understanding of the results obtained for OPSs in a general case of a noisy environment ($\bar{n}_{env}\neq0$), our subsequent discussion will initiate by focusing on an athermal environment ($\bar{n}_{env}=0$). This initial emphasis aims to demonstrate the evolution strategy of OPSs under athermal conditions. Drawing insights from these athermal cases, we will then explore the behavior of OPSs in a noisy environment.

\subsection{Athermal environment ($\bar{n}_{env}=0$)} \label{NFE}


Let's commence our discussion with the scenario of an athermal environment. In this case, the environmental state is designated as $\bar{n}_{env}=0$, corresponding to the noise state $\hat{\rho}_0=|0\rangle\langle0|$, representing a vacuum state. Under these conditions, we illustrate the advantages and characteristics of OPSs evolved through the optimization algorithm in Fig.\ref{Fig2}.

In Fig.\ref{Fig2} (a) and (b), we showcase the error probabilities for both coherent states and OPSs with a mean photon number of $\bar{n}=1$, along with the QA at different reflectivities. It is evident that OPSs exhibit a higher detection preference than coherent state probes across all reflectivity regions. In addition, two exciting phenomena are observed in the results. Firstly, an important observation is that the photon number distribution (the diagonal term of density matrix) of OPSs becomes concentrated as reflectivity increases, as shown in Fig.\ref{Fig2} (c)-(h). 

We first discuss the OPSs in the high-reflectivity region to understand the behavior. For the OPSs in QTD, it should be chosen to maximize the dissimilarity between $\hat{\rho}_1$ and $\hat{\rho}_0$. In the region of $r\rightarrow1$, $\hat{\rho}_1$ is primarily composed of the probe state $\hat{\rho}_{pr}$ with slight attenuation, while $\hat{\rho}_0$ represents a vacuum state. Therefore, the optimal strategy for constructing an OPS is to minimize the overlap with the vacuum state after reflection by the BS. An extreme case of $r=0.99$ is shown in Fig.\ref{Fig2} (h), for example, the mean photon number is set at $\bar{n}=1$. Consequently, the optimal strategy for OPS, in this case, is to concentrate the population as much as possible on the single-photon state, resulting in the OPS being in a single-photon Fock state. 

\begin{figure}[t]
\centering
\includegraphics[width=0.40\textwidth]{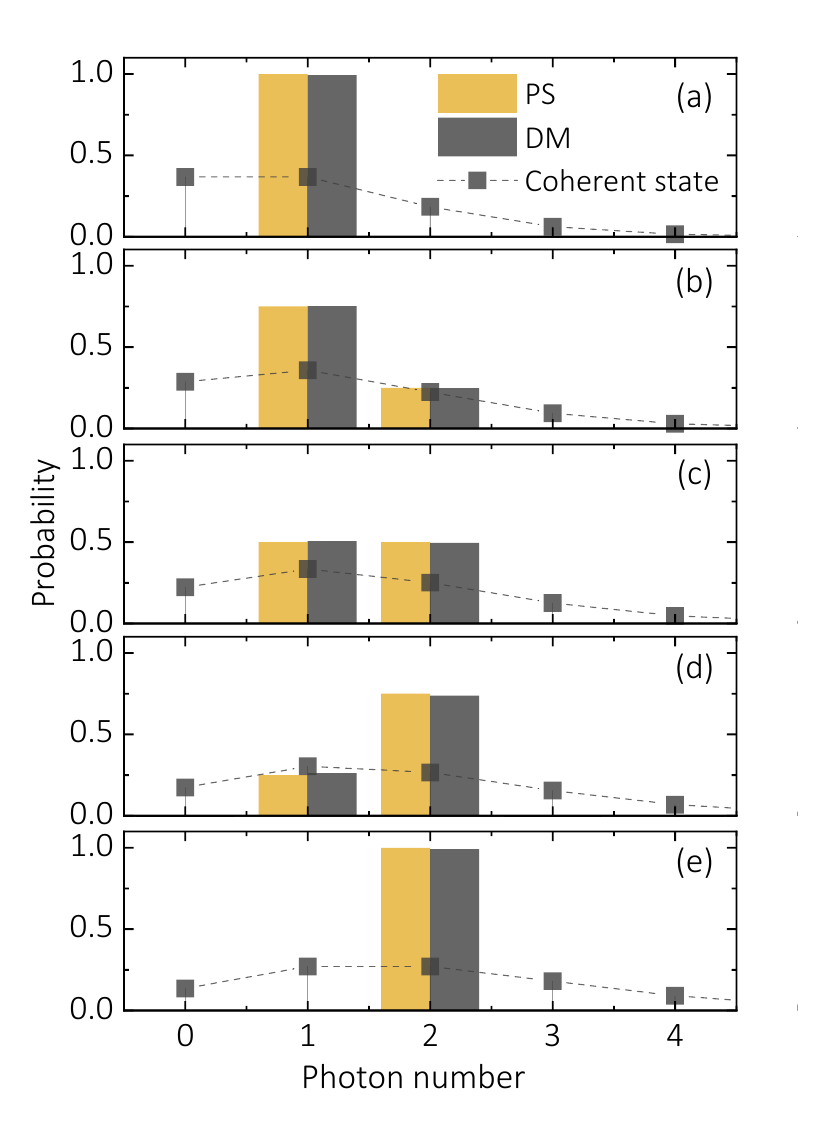}
\caption{Comparison of the diagonal term of OPSs obtained by the methods of photon statistics (PS, Eq.\ref{ps_eta0}) and density matrix (DM, Eq.\ref{P_err}).The mean photon number of probe state is set at (a) $\bar{n}=1$; (b) $\bar{n}=1.25$; (c) $\bar{n}=1.5$; (d) $\bar{n}=1.75$; (e) $\bar{n}=2$. The reflectivity is set at $r=0.99$ for all tests. The fidelity between the distributions calculated by the two methods is consistently higher than 0.999.
}
\label{Fig3}
\end{figure}

Expanding on the aforementioned concepts, intuitively, when considering the state constrained by Eq. \ref{n_bar}, any arbitrary combination of number states that avoids overlapping with $\left|0\right\rangle$ as much as possible will satisfy the conditions for OPSs as $r \rightarrow 1$, since $\hat{\rho}_{1}$ is quite close $\hat{\rho}_{pr}$. However, the inference above is incomplete; further exploration makes the results more inspiring and exciting. In Fig.\ref{Fig3}, we conducted tests with various $\bar{n}$ in the case of $r=0.99$ to study the behavior of OPSs in the high-reflectivity region, as represented by the gray bars. Notably, it is observed that the choice of $\bar{n}$ for OPSs is not arbitrary. OPSs tend to \emph{squeeze} the photon number distribution towards higher-number Fock states as much as possible. We term this phenomenon the photon-number squeezed states (PNSSs).

To understand the behavior, let's examine the characteristics of $\hat{\rho}_1$ in this high-reflectivity region. As $r\rightarrow1$, the state $\hat{\rho}_1$ is primarily influenced by $\hat{\rho}_{pr}$, with a slight attenuation caused by the imperfect reflectivity of the BS. After the attenuation, the photon number statistics undergo alterations, resulting in a redistribution of the population from higher photon-number states to lower photon-number states, thereby creating an overlap with the vacuum state. Ultimately, this gives rise to the error probability. To minimize this overlap, OPSs should concentrate the population as much as possible on higher photon number states. However, the energy limitation of Eq.\ref{n_bar} imposes a constraint on the highest occupation of the photon number state. Hence, OPSs are unable to distribute the population limitlessly across an infinitely high number of states, ultimately resulting in the emergence of PNSSs.

The core of the above idea is that, in the region where $r\rightarrow1$, the key information for distinguishing between the hypotheses of $H_0$ and $H_1$ primarily arises from the change in the diagonal term of the probe state's density matrix. In other words, if the calculation of error probability only considers the photon number statistics (the change in the diagonal term of the density matrix) but not the whole density matrix like Eq. \ref{P_err}, and further processes the optimal algorithm, the evaluated OPSs should still be the PNSSs in the $r\rightarrow1$ region.

To verify the concept, we assume that the impact of the non-unity reflectivity of the BS on $\hat{\rho}_{pr}$ can be analogized as a photon-number-counting measurement with non-unity efficiency. To capture this effect, we introduce photon number statistics with a finite efficiency \cite{scully1999quantum}
\begin{equation}
\begin{aligned}
P_{m}=\sum_{n=m}^{\infty}\binom{n}{m}r^{m}(1-r)^{n-m}\rho_{pr,nn}, 
\label{ps_eta}
\end{aligned}
\end{equation}
where $P_{m}$ is the probability that measured mm photon, $\rho_{pr,nn}$ is the photon number distribution of the initial probe state, and rr is the reflectivity or efficiency of photon-number-counting measurement. For $m=0$, it presents the probability of $\hat{\rho}_1$ occupying the vacuum state, which is contributed by the decayed nonzero photon states. Thus, we can obtain that from Eq.\ref{ps_eta} as
\begin{equation}
\begin{aligned}
P_{0}=\sum_{n=0}^{\infty}(1-r)^{n}\rho_{pr,nn}. 
\label{ps_eta0}
\end{aligned}
\end{equation}
As discussed above, $P_{0}$ is the source of error probability. Thus, OPSs should minimize $P_{0}$. Here, we implement the optimal algorithm on Eq.\ref{ps_eta0} again with different $\bar{n}$. The numerical results are shown by the yellow bars in Fig.\ref{Fig3}, demonstrating the good agreement between the two optimal methods, Eq.\ref{P_err} and Eq.\ref{ps_eta0}. It's important to emphasize that the methods of Eq.\ref{P_err} consider the whole density matrix (DM) to evaluate the OPSs, while, as a comparison, the methods of Eq.\ref{ps_eta0} only consider the diagonal terms (Photon statistics, PS). Based on the close results between these two methods, we conclude that OPSs don't require information about the phase (off-diagonal terms of the probe's density matrix) to discriminate between $H_0$ and $H_1$ in the $r\rightarrow1$ region. The majority of information is obtained from the photon number distribution. Furthermore, this implies that the optimal quantum measurement in the $r\rightarrow1$ region is the photon-number-counting measurement with a PNSS probe source. By observing the trend in Fig.\ref{Fig3}, we find the OPSs in $r\rightarrow1$, as
\begin{equation}
\left|\psi\right\rangle_{pr}^{opt} = \sqrt{p_n}\left|\left \lceil\bar{n}\right \rceil\right\rangle+\sqrt{1-p_n}\left|\left\lfloor\bar{n}\right\rfloor\right\rangle,
\label{PNSS}
\end{equation}
where $p_n=\bar{n}-\left \lceil\bar{n}\right \rceil+1$, $\left \lceil x \right \rceil=\min\{n\in\mathbb{Z}|x\geq n\}$ and $\left \lfloor x \right \rfloor=\min\{n\in\mathbb{Z}|x\leq n\}$.

\begin{figure*}[t]
\centering
\includegraphics[width=0.92\textwidth]{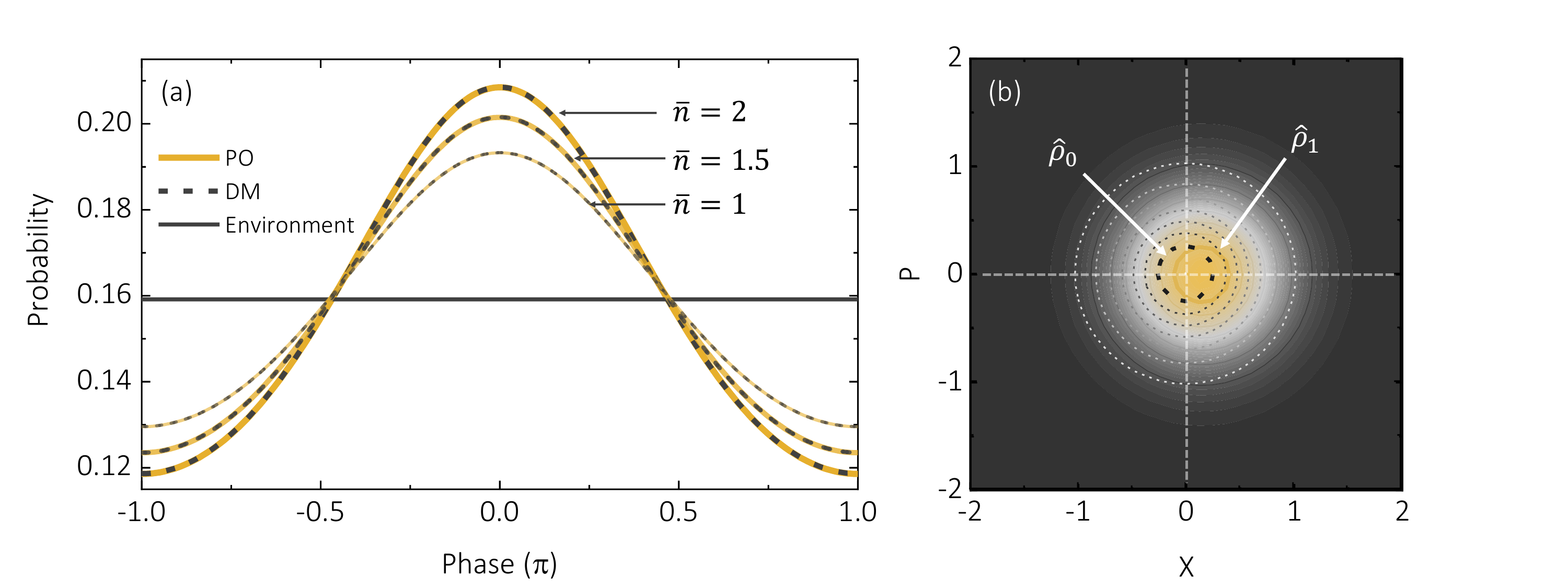}
\caption{(a) Comparison of phase distribution of OPSs obtained by the methods of phase overlapping (PO, Eq.\ref{phase_overlap}) and density matrix (DM, Eq.\ref{P_err}) in different mean photon numbers. The fidelity between the distributions calculated by the two methods is consistently higher than 0.999. (b) Demonstration of Wigner functions of vacuum state ($\rho_0$, dash line) and received state ($\rho_1$, color map) by using OPSs. The mean photon number of probe state is set at 1. The reflectivity is set at 0.01 for this figure. }
\label{Fig4}
\end{figure*}

Now, we discuss another region of low reflectivity. When $r\rightarrow0$, we can observe that the QA approaches 0, which means that the coherent state and OPSs now have the same error probabilities. To clarify the exact quantum state of OPSs in $r\rightarrow0$ regions, the fidelity between the coherent states and OPSs, 
\begin{equation}
F=\left[\text{Tr}\sqrt{\sqrt{\hat{\rho}_{coh}}\hat{\rho}_{opt}\sqrt{\hat{\rho}_{coh}}}\right]^2,
\label{Fidelity}
\end{equation}
is calculated and shown in Fig.\ref{Fig2} (b). The corresponding Wigner function and density matrix of OPS in the low-reflectivity case ($r=10^{-3}$) are also plotted in Fig.\ref{Fig2} (c) and (d), respectively. It is clear to see the OPSs in the $r\rightarrow0$ region tend to be in a coherent state since the fidelity approaches 1. The above results are consistent with and have been demonstrated in \cite{bradshaw2021optimal} using the method of Lagrange multipliers. 


Even though the fact that coherent states are the OPSs in the $r\rightarrow0$ region has been proven in \cite{bradshaw2021optimal}, here we present an alternative perspective on the physical insight of this result. Since the overlap between $\hat{\rho}_{recv}$ and $\hat{\rho}_{env}=|0\rangle\langle0|$ is unavoidable due to the high attenuation in the low reflectivity region, the optimal strategy to distinguish between $H_0$ and $H_1$ is to introduce the off-diagonal term of $\hat{\rho}_{pr}$ to induce coherence for $\hat{\rho}_{recv}$; in other words, the phase information of $\hat{\rho}_{recv}$ now contributes some information for distinguishing $H_0$ and $H_1$. Once a given $\hat{\rho}_{pr}$ enables a significant difference between the phase distribution of $\hat{\rho}_{recv}$ and $\hat{\rho}_{env}$, the given state is an OPS in the case of $r\rightarrow0$.

To describe the above idea quantitatively, here we introduce the phase distribution function, $P(\phi)$, for the quantum state $\hat{\rho}$, as \cite{loudon2000quantum}:
\begin{equation}
P(\phi)=\frac{1}{2\pi}\sum_{n,m=0}^{\infty}\rho_{n,m}e^{i(m-n)\phi},
\end{equation}
where $\rho_{n,m}$ is the density matrix element of the state $\hat{\rho}$, $\phi$ is the phase, and $P(\phi)$ is the probability of the phase distribution. It is easy to observe that when the state is a maximum mixed state or a number state (e.g., thermal state or vacuum state, $\rho_{n,m}=0$ for $n\neq m$), the phase distribution function is a constant of $1/2\pi$. This implies that those states have no defined phase information and are symmetrical relative to the origin point in the phase space. For OPSs in the low reflectivity region, efforts should be made to minimize the overlap in phase distribution of $\hat{\rho}_{recv}$ with the constant phase distribution of $1/2\pi$ in environment states. Thus, we proceed to calculate the overlap between these states, 
\begin{equation}
\langle P_{recv},P_{env}\rangle=\int_{-\pi}^{\pi}[P_{recv}(\phi)P_{env}(\phi)]^{\frac{1}{2}}d\phi,
\label{phase_overlap}
\end{equation}
where $P_{recv}(\phi)$ is the phase distribution of $\hat{\rho}_{recv}$ and $P_{recv}(\phi)$ is the phase distribution of $\hat{\rho}_{env}$. Eq.\ref{phase_overlap} now serves as a new target function for characterizing OPSs in the $r\rightarrow0$ region. Similar to the discussion in the high-reflectivity region, we investigate an alternative method by employing the optimal algorithm described in Eq.\ref{phase_overlap} to minimize the overlap and, consequently, identify the OPSs.

In Fig.\ref{Fig4}, we examine various $\bar{n}$ values for probe states, starting with an initial condition of $r=0.01$, to evaluate the optimal algorithm based on Eq.\ref{phase_overlap}. The resulting optimal phase distributions of OPSs obtained using this method are then presented. In Fig.\ref{Fig4} (a), it is evident that the phase distribution of the evaluated OPSs closely mirrors the distribution of coherent states used as probes. This observation implies that OPSs in the $r\rightarrow0$ region are near coherent states. Furthermore, since the optimal method in Fig.\ref{Fig4} is grounded in Eq.\ref{phase_overlap}, emphasizing our focus on utilizing phase distribution as the means to distinguish between $H_0$ and $H_1$, the results affirm our previously discussed idea—the primary information source is the phase of the received states in the $r\rightarrow0$ region. 

To highlight the effect of coherent states as OPSs in $r\rightarrow0$ regions, in Fig.\ref{Fig4} (b), we present the phase space representation of Wigner functions of received states with OPSs in both target-present and target-absent cases. It can be observed that after the injection of OPSs into the low-reflectivity BS, $\hat{\rho}_{recv}$ obtains a defined phase. Since the OPSs are the coherent states in this low-reflectivity case, the process can be likened to a displacement operator that acts on $\hat{\rho}_{env}$ \cite{paris1996displacement}, which maximally reduces the overlapping of Wigner functions between $\hat{\rho}_{env}$ and $\hat{\rho}_{recv}$.

Let's revisit Fig. \ref{Fig2} (b), where both fidelity and QA are shown as monotonic functions. This implies that OPSs translate monotonically from near-coherent states to PNSSs. Thus, OPSs in the whole reflectivity region are non-Gaussian states. On the other hand, the source of information for distinguishing between $H_0$ and $H_1$ gradually shifts from phase distribution to photon number statistics. Please note that the results mentioned above are for the cases where $n_{env}=0$. In the next section, we will start discussing the case where $\bar{n}_{env}\neq0$, which will reveal more interesting behaviors different from the case where $\bar{n}_{env}=0$.

\begin{figure}[t]
\centering
\includegraphics[width=0.4\textwidth]{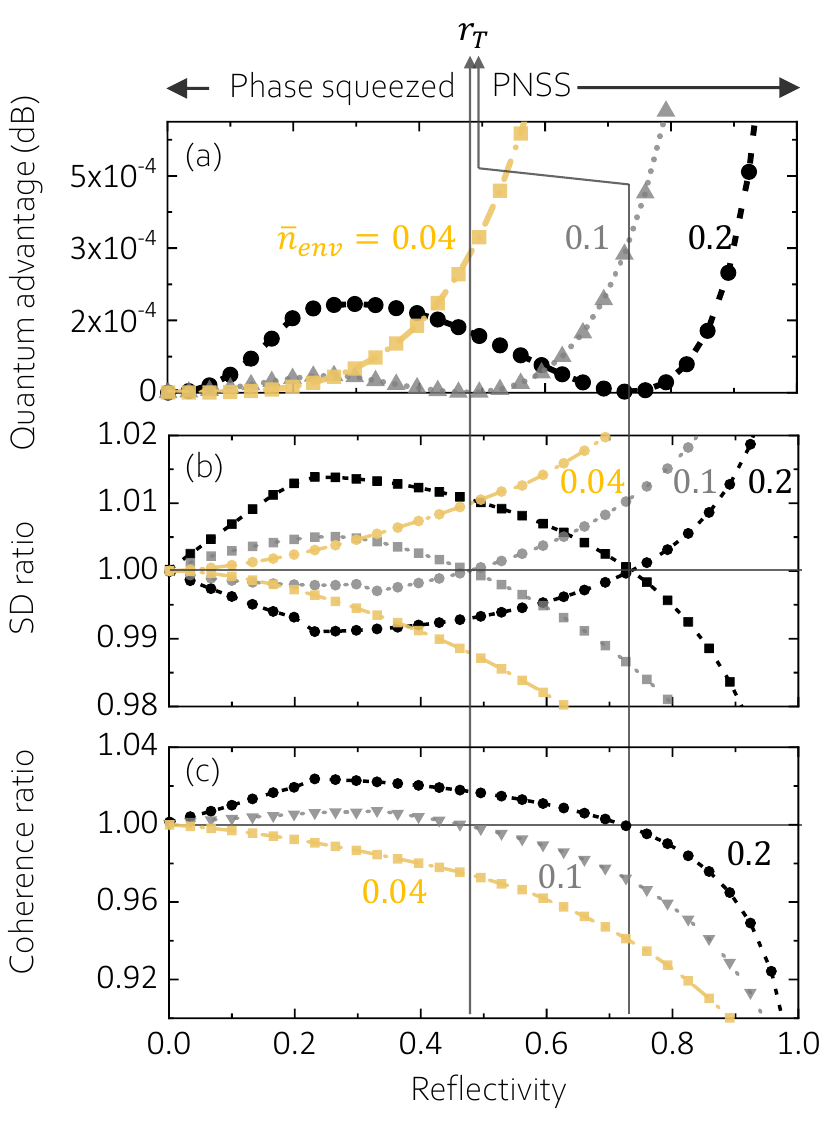}
\caption{(a) Quantum advantage of OPSs in the case of $\bar{n}_{env}=\{0.04;0.1;0.2\}$ which are represented by the color of yellow, gray, and black, respectively. The mean photon number is set at 0.04. (b) The standard deviation (SD) ratio of photon number (square) and phase (circle) between OPSs and coherent state is presented. (c) The coherence ratio between OPSs and coherent state with a series of $\bar{n}_{env}$ (as shown by the label in figure).
}
\label{Fig5}
\end{figure}

\subsection{Noisy environment ($\bar{n}_{env}\neq0$)}

In this section, we examine the scenario of a noisy environment, where $\bar{n}_{env} \neq 0$, and $\hat{\rho}_{env}$ represents a thermal state. Similar to the discussion in the preceding section, we initially compute the QA for various reflectivities of the BS, considering a range of $\bar{n}_{env}$ values, as depicted in Fig.\ref{Fig5} (a). Additionally, to comprehend the characteristics of OPSs in this context, we compute the variance of photon number distribution, $\Delta n^2_{OPS} = \bar{n^2} - \bar{n}^2$, and the full width at half maximum (FWHM) of the phase distribution, $\Delta \phi_{OPS}$, of the OPSs. To compare the differences between OPSs and coherent states, we further calculate the ratio between the photon number variance and the phase distribution of the two states (SD ratio). The SD ratio between these two states is illustrated in Fig.\ref{Fig5} (b).

In Fig.\ref{Fig5} (a), the QA approaches 0 in the $r\rightarrow0$ region. Moreover, in Fig.\ref{Fig5} (b), the ratio of the variance of the photon number distribution and the phase distribution both tends towards 1, indicating that OPSs behave approach coherent states in this scenario. This outcome mirrors the discussion in an athermal environment. In a noisy setting, the phase distribution of thermal states remains constant. Hence, the optimal discrimination method involves assigning a defined phase to $\hat{\rho}_{recv}$, while Eq. \ref{phase_overlap} can also be applied in this case. This results in OPSs behaving as coherent states even in the presence of noise in the $r\rightarrow0$ region.

In the $r\rightarrow1$ regions, as depicted in Fig.\ref{Fig5} (b), the photon number variance of OPSs gradually decreases, indicating that OPSs are evaluated towards PNSSs. Similar to the discussion in an athermal environment, in the noisy and $r\rightarrow1$ region, the probe states now dominate the received states but with a slight leakage of thermal states. As thermal states exhibit a decreasing photon number distribution, the optimal strategy for constructing OPSs is to allocate the population to higher photon number states to minimize the overlap between received and thermal states. Consequently, OPSs manifest as PNSSs in noisy environments and as $r\rightarrow1$.

Unlike the athermal case, however, an interesting behavior is observed in Fig. \ref{Fig5}, where the QA and SD ratio do not always evolve monotonically and depend on the value of $\bar{n}_{env}$ for those cases. According to the simulation, the QA and SD ratio exhibit monotonic behavior for approximately $\bar{n}_{env} \leq \bar{n}$. In this case, since $\bar{n}_{env} \leq \bar{n}$, the received states are consistently dominated by $\hat{\rho}_{pr}$; hence, the behaviors of QA and OPS align with those observed in the athermal environment, as discussed in Sec. \ref{NFE}. Similar results are illustrated by the SD ratio of $\bar{n}_{env}=0.04$ in Fig. \ref{Fig5} (b). It can be observed that both the $\Delta n^2_{OPS}$ and $\Delta \phi_{OPS}$ of OPSs are decreasing and increasing, respectively, implying that the OPSs are evolving into PNSSs throughout the entire reflectivity region.

In the case of $\bar{n}_{env} > \bar{n}$, however, the trend of QA and SD ratio does not exhibit monotonic behavior. In Fig. \ref{Fig5} (a), we can observe that QA has an inflection point at a specific reflectivity, $r_T$; furthermore, the QA is close to 0 when $r=r_T$. In Fig. \ref{Fig5} (b), the SD ratio also exhibits a non-monotonic behavior. It is noteworthy that the SD ratio is equal to 1 when $r=r_T$. Both results indicate that not only the coherent state is the OPSs for $r\rightarrow0$, but coherent states also serve as the OPSs for $r=r_T$. 

Another crucial discovery is that when $0<r<r_T$, the SD ratio indicates that OPSs have a more precise phase distribution than coherent states (and higher variance in the photon number distribution), implying that OPSs are phase-squeezed states within this region. To comprehend this characterization, one should realize that this situation is quite different from the case of an athermal environment. For $\bar{n}_{env} > \bar{n}$, the $\hat{\rho}_{env}$ now contributes unavoidable noise photons on the diagonal terms of the state $\hat{\rho}_{recv}$ not only in the region of $r\rightarrow0$ but also at a finite reflectivity to a certain degree. Thus, the OPSs cannot always follow the strategy used in the athermal case (PNSSs as probe states). To optimally distinguish between $\hat{\rho}_{env}$ and $\hat{\rho}_{recv}$, the off-diagonal terms (the coherence) of the probe states now play a crucial role in providing the difference between $\hat{\rho}_{env}$ and $\hat{\rho}_{recv}$, further resulting in the phase-squeezed state in the $0<r<r_T$ region. 

To validate the idea, we introduce the size of off-diagonal elements to quantify the coherence, denoted as $\mathcal{C}$, of the OPSs and further analyze its relation with the reflectivities. The coherence of the quantum state $\hat{\rho}$ is defined as \cite{baumgratz2014quantifying}, 
\begin{equation}
\mathcal{C}(\hat{\rho})=\sum_{n\neq m}^{\infty}|\rho_{n,m}|,
\end{equation}
where $\rho_{n,m}$ is the off-diagonal matrix elements of $\hat{\rho}$. In Fig. \ref{Fig5} (c), we depict the coherence ratio between OPSs and coherent states with a fixed mean photon number of $\bar{n}=0.04$. The trend of the coherence ratio illustrates that under the conditions of $0<r<r_T$ and $\bar{n}_{env} > \bar{n}$, the coherence of OPSs surpasses that of coherent states. This observation supports our perspective that coherence plays a crucial role in distinguishing between $\hat{\rho}_{env}$ and $\hat{\rho}_{recv}$ within the conditions of $0<r<r_T$ and $\bar{n}_{env} > \bar{n}$.

We now elucidate and consolidate the behaviors and properties of OPSs across the entire range of $r$. In Fig.\ref{Fig5}, in the region $0<r<r_T$, as $r$ increases, the contribution of the photon number distribution from the thermal states gradually loses influence on $\hat{\rho}_{recv}$. Thus, the optimal strategy for distinguishing between $\hat{\rho}_{recv}$ and $\hat{\rho}_{env}$ starts to shift from mainly using coherence to a combination of both photon number distribution and coherence. This results in the SD ratio of OPSs gradually tending towards 1. Finally, when $r=r_T$, the OPSs revert to coherent states. When $r>r_T$, $\hat{\rho}_{recv}$ start to be dominated by $\hat{\rho}_{pr}$. Therefore, the behavior of OPSs tends to PNSSs, as demonstrated in Fig. \ref{Fig5} (b). 

\begin{figure*}[htbp]
\centering
\includegraphics[width=.75\textwidth]{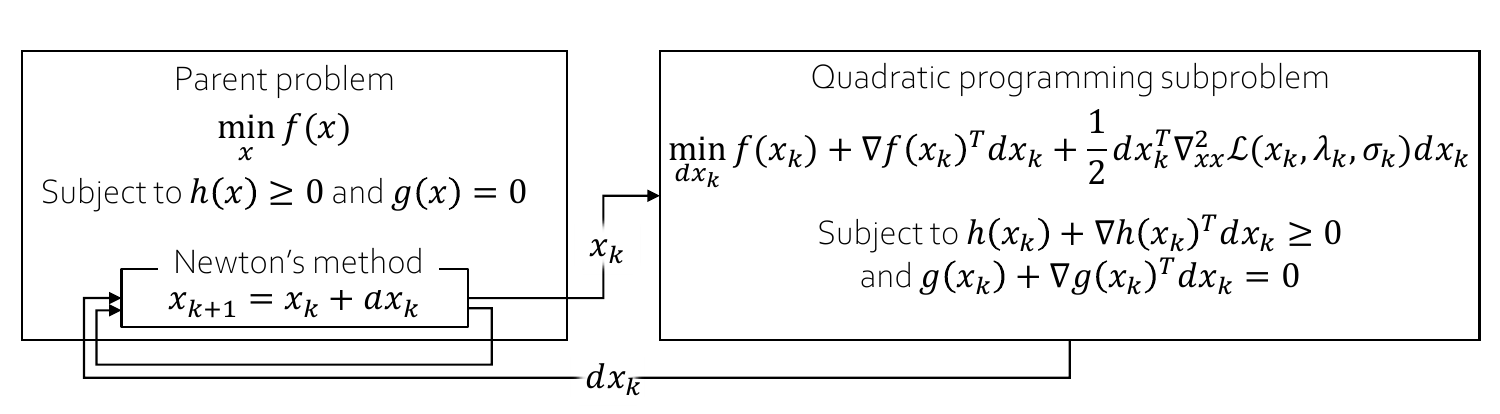}
\caption{Schematic diagram for the basic SQP algorithm.
$\mathcal{L}$ is the Lagrangian for the problem, $\lambda_k$ and $\sigma_k$ are Lagrange multipliers. $\nabla^{2}_{xx}$ is Hessian matrix.}
\label{SQP}
\end{figure*}

\section{Summary} \label{secIV}

In this section, we consolidate the study and results from the preceding sections. Based on the preceding discussion, we can distill OPSs into two scenarios. In the case where $\bar{n}\geq\bar{n}_{env}$, the characteristics of $\hat{\rho}_{pr}$ wield a dominant influence on the received states throughout the entire range of $r$. Consequently, OPSs consistently shift towards PNSS as $r$ increases. Conversely, when $\bar{n}<\bar{n}_{env}$, OPSs display features that allow us to classify them into two distinct regions. For $r<r_T$, the coherence of OPSs surpasses that of coherent states, presenting phase-squeezed states that predominantly utilize the off-diagonal term of $\hat{\rho}_{pr}$ as the information source for executing QTD. When $r>r_T$, $\hat{\rho}_{pr}$ recapture the domination of $\hat{\rho}_{recv}$, the OPSs turn back to PNSSs. In addition, regardless of whether $\bar{n}\geq\bar{n}_{env}$ or $\bar{n}<\bar{n}_{env}$, the OPSs are near-coherent states and PNSSs (Eq.\ref{PNSS}) as $r$ tends towards 0 and 1, respectively.

At the specific point $r=r_T$, identified as a transition point, OPSs transition from phase-squeezed states to PNSSs. In other words, $r=r_T$ marks a demarcation point for the primary information utilized in QTD. When $r<r_T$, the off-diagonal term of OPSs takes main effect in distinguishing between the $H_0$ and $H_1$. For $r>r_T$, it transitions to the diagonal term of OPSs.

Intuitively, as the ratio of $\bar{n}_{env}/\bar{n}$ increases with the growth of $r_T$, it implies that $\hat{\rho}_{recv}$ requires a greater composition of probe states to counteract the influence of $\hat{\rho}_{env}$. This speculation is also illustrated in Fig.\ref{Fig5}. Interestingly, OPSs assume the form of coherent states at $r=r_T$, resulting in the absence of quantum advantage at this specific point. This occurrence presents a potential strategy for the target or jammer to weaken the impact of QTD.

\section{Conclusion}\label{secV}

In conclusion, this study utilizes an optimization algorithm to identify single-mode OPSs across the entire range of target reflectivity, thereby complementing the existing knowledge of OPSs in non-extreme detection conditions. Additionally, through a comparative analysis with alternative methods for OPSs assessment, we provide a clear physical interpretation of the observed phenomena. For future work, expanding the analytical approach to include two- or even higher-mode QTD could facilitate the identification of OPSs with entanglement and contribute to a more comprehensive theoretical model of quantum illumination and quantum target detection. On the other hand, from the perspective of the target and jammer, their objective is to diminish the performance of the prober's quantum advantage. In this scenario, the jammer could emit various possible noise quantum states arbitrarily to reduce the information obtained by the prober. Therefore, from the perspective of both the target and the jammer, identifying an optimal jammer (noise) quantum state will also be a crucial future task in defending against quantum target detection.

\section*{Acknowledgements}
This work was supported by the National Science and Technology Council (NSTC) of Taiwan under Grants 112-2112-M-008-025-MY3. We also thank Dr. Kuan-Ting Lin at National Taiwan University for his fruitful discussions.

P.-J. T. and W.-M. C. contributed equally to this work.

\section*{Appendix}

In this section, we introduce the optimal algorithm, sequential quadratic programming (SQP), used to evaluate the OPSs in this study. SQP is a numerical optimization algorithm utilized to minimize a nonlinear objective function while adhering to equality and inequality constraints. 

Fig.\ref{SQP} presents the flow chart of SQP. Consider a parent problem that needs to find the solution $x_k$ to minimize a target function $f(x)$. Newton's method provides an iterative approach for finding this solution. However, when the problem includes constraints such as $h(x) \geq 0$ and $g(x) = 0$, SQP suggests considering a quadratic programming (QP) subproblem to calculate the Newton step direction $dx_k$ in order to generate a better approximation $x_{k+1}$. In QP subproblem, the target function is transformed to 
\begin{equation} 
     f(x_{k}) + \nabla f(x_{k})^T dx_{k}+\frac{1}{2}dx_{k}^{T} \nabla_{xx}^{2} \mathcal{L}(x_{k},\lambda_k,\sigma_k)dx_{k},
\end{equation}
with the constraints that be considered to the first-order term of Taylor expansion
\begin{equation} 
\begin{aligned}
   h(x_{k}) + \nabla h(x_{k})^T dx_{k} &\geq 0, \\
   g(x_{k}) + \nabla g(x_{k})^T dx_{k} &= 0.
\end{aligned}
\end{equation}

We begin the algorithm by initializing a set of parameters, denoted as $[x_0,\lambda_0,\sigma_0]$. These initial values are then fed into the QP subproblem to compute the first Newton step, $dx_{0}$, using the Lagrange multiplier method. Subsequently, the parameter set in the parent problem is updated as $[x_1,\lambda_1,\sigma_1]^T=[x_0,\lambda_0,\sigma_0]^T+dx_0$. This updated parameter set serves as the initial condition for the second round of the QP subproblem and is utilized to compute the subsequent Newton step, $dx_{1}$. This process is repeated iteratively, updating the parameter set as $[x_{k+1},\lambda_{k+1},\sigma_{k+1}]^T=[x_k,\lambda_k,\sigma_k]^T+dx_k$, until the parent problem converges. Finally, the resulting $x_{k}$ values that minimize $f(x)$ are obtained, providing the solution to the optimization problem.

In our case, the target function $f(x)$ is the error probability (Eq.\ref{P_err}), while the constraint $g(x)$ is defined by the normalization condition, Eq.\ref{nor}, and the finite mean photon number of probe states, Eq.\ref{n_bar}. Due to computational limitations, the dimension of the calculated density matrix is restricted to 8, spanning from the vacuum state $|0\rangle$ to the number state $|7\rangle$. By evaluating the algorithm with these equations and conditions, we can determine the optimal probe states under arbitrary detection conditions.

\bibliography{refs}
\end{document}